# *Contactless Rheology of Soft Gels over a Broad Frequency Range*

*Zaicheng Zhang[1], Muhammad Arshad[1], Vincent Bertin[1,,2,3], Samir Almohamad[1,4], Elie Raphaël[2], Thomas Salez[1] and Abdelhamid Maali[1, *]*

[1] *Univ. Bordeaux, CNRS, LOMA, UMR 5798, F-33405, Talence, France.*
[2] *UMR CNRS Gulliver 7083, ESPCI Paris, PSL Research University, 75005 Paris, France.*
[3] *Physics of Fluids Group, Faculty of Science and Technology, Mesa+ Institute, University of Twente, 7500 AE Enschede, The Netherlands.*
[4] *Univ. Lille, CNRS, Centrale Lille, Yncra ISEN, Univ. Polytechnique Hauts-de-France, UMR 8520 – IEMN, F-59000 Lille, France.*

**Abstract:**
*We report contactless measurements of the viscoelastic rheological properties of soft gels. The experiments are performed using a colloidal-probe Atomic Force Microscope (AFM) in a liquid environment and in dynamic mode. The mechanical response is measured as a function of the liquid gap thickness for different oscillation frequencies. Our measurements reveal an elastohydrodynamic (EHD) coupling between the flow induced by the probe oscillation and the viscoelastic deformation of the gels. The data are quantitatively described by a viscoelastic lubrication model. The frequency-dependent storage and loss moduli of the polydimethylsiloxane (PDMS) gels are extracted from fits of the data to the model and are in good agreement with the Chasset–Thirion law. Our results demonstrate that contactless colloidal-probe methods are powerful tools that can be used for probing soft interfaces finely over a wide range of frequencies.*

\* Corresponding author: abdelhamid.maali@u-bordeaux.fr

*Introduction:*

Mechanical properties of thin, soft solids are of great interest for many emerging applications including surface coatings [1-3], photonics [4,5], microelectronics [6,7], and biosensors [8-10]. Methods like tensile or bending tests and direct indentation are commonly used for the characterization of soft materials [11-14]. The Atomic Force Microscope (AFM) is a widely spread device for surface topography and is also used for mechanical measurements with a nanometric resolution. AFM probes, with known shapes, can be used as nano-indenters to measure the mechanical response of thin films under applied normal compression [15-23]. With such a device, one can perform nanomechanical characterization of materials with elastic moduli in the range from $kPa$ to $GPa$.

However, there are limitations in the measurement of elastic moduli for soft samples using the direct contact method due to the adhesion between the sample and the probe, which is ineluctable at the nanoscale [23-26]. It is not only difficult to accurately measure adhesion, but also to measure the area of an adhesive contact [12,27,28]. Sample damage and probe contamination may also occur during the measurements on soft samples [29].

Contactless measurements are an alternative to direct-contact methods. More precisely, elastohydrodynamic (EHD) interactions between a probing sphere and a soft sample in a liquid environment provide a new and precise method to assess mechanical properties [30,31]. At small liquid-gap thickness between the sphere and the soft substrate, the vibration of the former generates a hydrodynamic stress field that might be large enough to deform the substrate, which in turn perturbs the flow leading to an EHD coupling. Based on this coupling, new tools were developed to probe the mechanical properties of soft interfaces without contact. Using surface force apparatus (SFA), the mechanical properties of substrates of various rigidities, ranging from soft elastomeric samples to hard glasses, have been measured at low frequencies (less than 100 Hz) [32,33]. Guan et al. have used a vibrating nano-needle glued onto an AFM cantilever to probe the viscoelastic properties of polydimethylsiloxane (PDMS) samples [34] and living cells [35]. However, in the AFM measurements of Guan et al. [34,35], the experimental setup restricts the probed frequency to a single value given by the resonance of the cantilever.

Usually, the frequency-dependent rheology of soft solids is probed at the macroscale with conventional shear rheometers, where the sample is fixed between two rotating parallel plates [36-39]. Imposing a rotational displacement to one plate and measuring the applied torque on the other plate allows one to extract the complex shear modulus of the system, $\mu(\omega) = \mu'(\omega) + j\mu''(\omega)$, where $\omega$ is the angular frequency, $\mu'(\omega)$ is the storage shear modulus, $\mu''(\omega)$ is the loss shear modulus [40-42], and $j$ is the imaginary number such that $j^2 = -1$. This method requires perfect contact between the sample and the plates of the rheometer. In contrast, the dynamic colloidal AFM method [43] is a good candidate to probe the substrate rheology at the nanoscale for different frequencies and without such a constraint. It offers the key advantage of a large frequency range using the same colloidal probe, from 20 Hz up to a few kHz.

In this Letter, we present measurements of the mechanical response of an immersed, thin, soft polymeric gels using an AFM colloidal probe, as a function of the separation distance between the probe and the soft substrate. From the mechanical impedance, we obtain simultaneously the loss and storage moduli of the soft solid. The measured viscoelastic properties of the thin PDMS samples at different frequencies are in good agreement with the *Chasset–Thirion* model. Our study demonstrates that contactless colloidal probes are a powerful tool to characterize soft interfaces in a non-invasive manner and over a large range of frequencies.

*Sample preparation:*

The studied samples are PDMS substrates prepared as follows. First, uncrosslinked PDMS (Sylgard 184, Dow Corning) and a curing agent are mixed with a weight ratio of 70:1. Following a degassing process in vacuum, a droplet of the mixture is spin-coated on a cover slide with a size of *24* mm × *24* mm for a minute, to get a sample with thickness $e = 26 \pm 3$ μm. The sample is then annealed in the oven at 50 °C and for 24 h, to promote an efficient cross-linking. The employed viscous liquid is dodecane with a viscosity of $\eta = 1.34$ mPa · s. After swelling, the measured thickness of the sample is $125 \pm 5$ μm.

*Experimental setup and methods:*

The experiments are performed using an AFM (Bioscope, Brucker) equipped with a liquid cell (DTFML-DDHE) that allows one to perform measurements in a liquid medium and that provides a clean cantilever excitation spectrum (free from spurious peaks). The schematic of the experimental setup is shown in Fig. 1. A borosilicate sphere with a radius $R = 55 \pm 1$ μm is glued at the end of a silicon nitride cantilever (SNL-10, Brucker), and the stiffness of the cantilever is determined by the drainage method, to be $k_c$=0.25 ± 0.02 N/m [44]. The resonance frequency and bulk quality factor are obtained from thermal spectral density as $\omega_0/(2\pi) = 1375 \pm 5$ Hz and $Q = 2.8 \pm 0.1$, respectively. The sample is fixed on a piezo stage (NanoT series, Mad City Labs), which is used to control the distance between the sphere and the sample by imposing a displacement to the substrate at low velocity (less than 160 nm/s). The cantilever is excited by the base oscillation $A_b e^{j\omega t}$ [43,45], where $A_b$ and $\omega$ are the amplitude and angular frequency of the base oscillation, respectively, and where $t$ denotes times. The amplitude $A$ and phase $\varphi$ of the oscillation of the cantilever are measured by a lock-in amplifier (Model 7280, signal recovery) versus the piezo displacement. In addition, the DC component of the cantilever's deflection is also recorded to determine the distance between the sphere and the sample's surface.

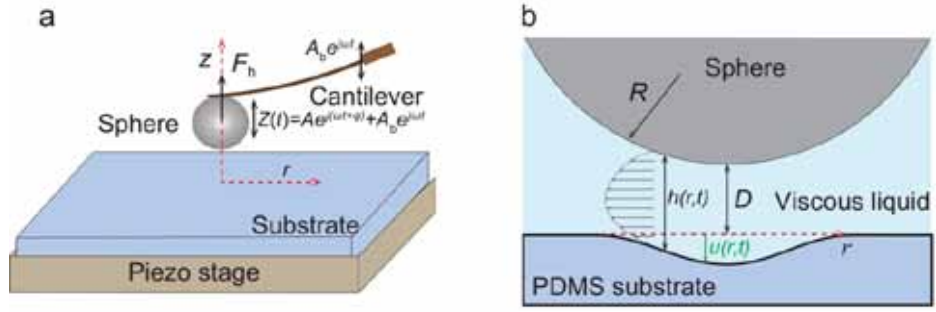

*Figure 1: a) Schematic of the experimental setup. A borosilicate sphere is glued on the edge of an AFM cantilever, which is vibrated vertically near the soft PDMS substrate within the viscous liquid (dodecane, $\eta = 1.34$ mPa · s ) by the base oscillation $A_b e^{j\omega t}$. b) The motion of the sphere gives rise to an axisymmetric liquid-gap thickness profile $h(r,t)$ depending on the radial position $r$ and time $t$, and related to the sample deformation $u(r,t)$, the average distance D and the sphere vibration $Z(t)$.*

The system is modeled as a forced damped harmonic oscillator, and the vertical displacement $Z(t)$ of the sphere satisfies [43,45]:
$$m^* \ddot{Z} + \gamma_b \dot{Z} + k_c Z = F_d + F_h, \quad (1)$$

where $m^*$ is the effective mass of the probe, $\gamma_b$ is the damping coefficient in the bulk liquid, $k_c$ is the cantilever's stiffness, $F_d$ is the driving force acting on the cantilever and $F_h$ is the EHD force induced by the flow between the sphere and the deformable sample. Using the Ansatz $Z(t) = A e^{j(\omega t + \varphi)} + A_b e^{j\omega t}$, the complex mechanical impedance $G = G' + jG''$ can be defined and determined by [43,46]:

$$G = -\frac{F_h}{Z} = -k_c \left[1 - \left(\frac{\omega}{\omega_0}\right)^2 + j\frac{\omega}{\omega_0 Q}\right] \frac{A e^{j\varphi} - A_\infty e^{j\varphi_\infty}}{A e^{j\varphi} + A_b}, \quad (2)$$

where $A_\infty$ and $\varphi_\infty$ are respectively the amplitude and phase measured far from the surface, *i.e.* where the EHD force vanishes.

Figure 2 shows typical acquired data. Figure 2a presents the amplitude and phase ($A_\infty, \varphi_\infty$) versus the oscillation frequency, far from the surface ($D > 400$ μm), where the EHD interaction between the substrate and the sphere can be neglected. Figure 2b presents the measured amplitude A (red dots) and phase $\varphi$ (green dots) as functions of the average separation distance D, for an oscillation frequency $\omega/(2\pi) = 500$ Hz. Additional measurements (blue dots and black dots respectively) on a hard surface (silicon) are used to obtain the base oscillation $A_b$. Indeed, the silicon surface is hard enough for us to neglect the sample deformation. Thus, in that case, the interaction is purely viscous and can be modeled by a Reynolds force ($G' = 0, G'' = 6\pi\eta R^2 \omega/D,$) [43,47]. Very close to the silicon surface, the imaginary component $G''$ of the impedance tends to infinity ($D \to 0, G'' \to \infty$), and the amplitude (blue dots in Fig. 2b) of the cantilever oscillation tends to the value of the base oscillation ($A \to A_b$) [43]. In practice, in Fig. 2(b), the amplitude of base oscillation is determined as $A_b = 3.1$ nm at $\omega/(2\pi) = 500$ Hz. Once this parameter is measured, the complex mechanical impedance $G$ can be computed using Eq. (2).

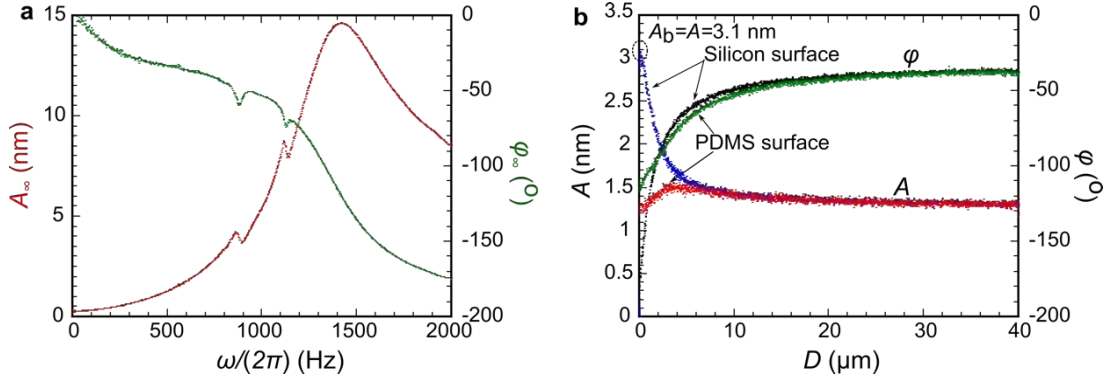

*Figure 2: a) Amplitude and phase of the cantilever's oscillation versus oscillation frequency measured far from the surface (D > 400 μm). b) Amplitude and phase measured as functions of the average separation distance between the sphere and either a soft PDMS substrate or a rigid silicon substrate. When the sphere is very close to the silicon surface, the amplitude A is equal to the base amplitude $A_b$ =3.1 nm.*

*Elastohydrodynamic model:*

We focus on the small-gap regime, *i.e.* $D \ll R$, and we adopt the lubrication approximation. The no-slip boundary condition is assumed at both the sphere and substrate surfaces, so that the liquid-gap thickness $h(r,t)$ obeys the Reynolds equation [47]:

$$\frac{\partial h(r,\ t)}{\partial t} = \frac{1}{12\eta r}\frac{\partial}{\partial r}\left(rh(r,\ t)^3 \frac{\partial}{\partial r}p_e(r,\ t)\right), \tag{3}$$

where $p_e(r,\ t)$ is the excess hydrodynamic pressure with respect to the atmospheric pressure, with $r$ and $t$ the radial coordinate and time, respectively. The liquid-gap thickness (see Fig. 1b) is related to the substrate's deformation $u(r,t)$ and the sphere's vibration amplitude $Z(t)$, through [47]:

$$h(r,\ t) = D + \frac{r^2}{2R} + Z(t) + u(r,\ t). \tag{4}$$

The amplitude of the base oscillation and the sample deformation are very small compared to the average gap distance *D*, and thus Eq. (3) can be linearized to get:

$$\frac{dZ(t)}{dt} + \frac{\partial u(r,\ t)}{\partial t} = \frac{1}{12\eta r}\frac{\partial}{\partial r}\left(r\left(D + \frac{r^2}{2R}\right)^3 \frac{\partial}{\partial r}p_e(r,t)\right). \tag{5}$$

The cantilever's motion, the substrate's deformation and the excess pressure are assumed to have harmonic time dependences, in the form: $Z(t) = Z_0 e^{j\omega t}$, $u(r,\ t) = U(r)e^{j\omega t}$ and $p_e(r,\ t) = P(r)e^{j\omega t}$, respectively. Equation (5) then becomes:

$$12j\eta r\omega(Z_0 + U(r)) = \frac{d}{dr}\left(r\left(D + \frac{r^2}{2R}\right)^3 \frac{d}{dr}P(r)\right). \tag{6}$$

Integrating Eq. (6) once, we get [47]:

$$\frac{d}{dr}P(r) = \frac{6j\eta\omega r\, Z_0}{\left(D + \frac{r^2}{2R}\right)^3} + \frac{12j\eta\omega}{r\left(D + \frac{r^2}{2R}\right)^3}\int_0^r r'U(r')\, dr'. \tag{7}$$

The substrate is assumed to be a semi-infinite viscoelastic medium with Poisson ratio $\nu$ and complex Young's modulus $E(\omega)$, including the frequency-dependent storage modulus $E'(\omega)$ (real part) and loss modulus $E''(\omega)$ (imaginary part). Therefore, the deformation is related to the pressure field through [47,48]:

$$\widehat{U}(k) = \frac{2\widehat{P}(k)}{E^*(\omega)\, k}, \tag{8}$$

where $E^*(\omega) = E(\omega)/(1-\nu^2)$ is the reduced Young's modulus, and where $\widehat{U}(k)$ and $\widehat{P}(k)$ are the zeroth-order Hankel transforms of the deformation $U(r)$ and the pressure $P(r)$, respectively, defined as: $\widehat{U}(k) = \int_0^{+\infty} J_0(kr)U(r)r\, dr$ and $\widehat{P}(k) = \int_0^{+\infty} J_0(kr)P(r)r\, dr$, with $J_0$ the zeroth-order Bessel function of the first kind. For incompressible materials, $\nu = 0.5$ and thus $E^* = \frac{4}{3}E$. In such a case, and inverting the transformation, we get:

$$U(r) = \int_0^\infty J_0(kr)\widehat{U}(k)k\, dk = \int_0^\infty J_0(kr)\frac{3\widehat{P}(k)}{2E(\omega)}\, dk. \tag{9}$$

Injecting Eq. (9) into Eq. (7) and using the relation between the zeroth-order and first-order ($J_1$) Bessel functions of the first kind, i.e. $\int_0^r J_0(kr')\, r'\, dr' = rJ_1(kr)/k$, we get:

$$\frac{d}{dr}P(r) = \frac{6j\eta\omega r Z_0}{\left(D + \frac{r^2}{2R}\right)^3} + \frac{18j\eta\omega}{\left(D + \frac{r^2}{2R}\right)^3 E(\omega)}\int_0^\infty \frac{J_1(kr)\,\widehat{P}(k)}{k}\, dk. \tag{10}$$

We further introduce the dimensionless variables:

$$x = \frac{r}{\sqrt{2RD}},\; q = k\sqrt{(2RD)},\; p = \frac{PD^2}{\eta R Z_0 \omega},\; D_c = 8R\left(\frac{3\eta\omega}{4E'(\omega)}\right)^{\frac{2}{3}},\; B(\omega) = \frac{E''(\omega)}{E'(\omega)}.$$

Using these variables, Eq. (10) becomes:

$$\frac{d}{dx}p(x) = \frac{12jx}{(1+x^2)^3} + \frac{3j}{(1+x^2)^3}\left(\frac{D_c}{D}\right)^{\frac{3}{2}}\frac{1-jB}{1+B^2}\int_0^{+\infty}\frac{\widehat{p}(q)J_1(qx)}{q}\, dq, \tag{11}$$

where $\widehat{p}(q)$ is the zeroth-order Hankel transform of $p(x)$. A first-order Hankel transform of Eq. (11) leads to a Fredholm integral equation for $\widehat{p}(q)$:

$$\widehat{p}(q) = \frac{3j}{2}qK_1(q) + 3j\left(\frac{D_c}{D}\right)^{\frac{3}{2}}\frac{1-jB}{1+B^2}\int_0^\infty \widehat{p}(q')\, dq'\int_0^{+\infty}\frac{J_1(q'x)J_1(qx)}{(1+x^2)^3 qq'}\, x\, dx, \tag{12}$$

where $K_1$ is the first-order modified Bessel function of the second kind. The mechanical impedance is then given by [47]:

$$G(D) = -\frac{1}{Z_0}\int_0^{+\infty} 2\pi\, rP(r)dr = -\frac{4\pi\eta R^2\omega}{D}\widehat{p}(0), \tag{13}$$

which can be rescaled as:

$$G(D) = \frac{6\pi\eta R^2 \omega}{D_c} g_c\left(\frac{D}{D_c}, B\right), \qquad g_c\left(\frac{D}{D_c}, B\right) = -\frac{4D_c}{6D} \hat{p}(0). \qquad (14)$$

The dimensionless function $g_c\left(\frac{D}{D_c}, B\right)$ can be computed numerically from Eq. (12) [47]. Doing so, we obtain the mechanical impedance. In particular, at large distance, the mechanical response is dominated by the viscous contribution; hence, we get the asymptotic expression of the impedance:

$$G(D) \simeq \frac{6\pi\eta R^2 \omega}{D}\left(j + \frac{9\pi^2}{512}\left(\frac{D_c}{D}\right)^{3/2} \frac{1}{1+B^2}\right). \qquad (15)$$

***Results and discussion:***

Figure 3 shows the dimensionless mechanical impedance $GD_c/(6\pi\eta R^2\omega)$ as a function of the dimensionless distance $D/D_c$, for two frequencies. Figure 3a shows the results obtained from the data shown in Fig. 2b and using Eq. (2), where the oscillation frequency is $\omega/(2\pi) = $ *500* Hz. The solid black lines correspond to Eq. (14), where the values of the storage modulus $E' = 3.2 \pm 0.3$ kPa and loss modulus $E'' = 2.1 \pm 0.2$ kPa are the fit parameters. Figure 3b shows the results for the frequency $\omega/(2\pi) = 50$ Hz. Similarly, we obtain the values of the storage modulus $E' = 1.6 \pm 0.2$ kPa and loss modulus $E'' = 0.6 \pm 0.1$ kPa.

Furthermore, at large distances ($D > D_c$), the viscous component $G''$ of the mechanical impedance dominates and follows a $\sim D^{-1}$ scaling law (see Eq. (15)), corresponding to the asymptotic expression of the hydrodynamic damping between a sphere and a rigid plane [47]: $G''(D) = 6\pi\eta R^2\omega/D$. Nevertheless, the deformation of the substrate leads to a non-zero elastic component $G'$ of the mechanical impedance, that follows a $\sim D^{-5/2}$ scaling law, in good agreement with the asymptotic prediction of the viscoelastic lubrication model (see Eq. (15)).

At small distances ($D < D_c$), both $G'$ and $G''$ saturate to constant values that are independent of the average gap thickness. This is due to the fact that the deformation of the soft sample cannot exceed the oscillation amplitude of the sphere, and thus saturates, leading to a saturation of the excess pressure. In this near-contact regime, the sample deformation accommodates the sphere motion, and the liquid is no longer expelled from the gap. Besides a residual damping that occurs outside the gap due to the viscous flow therein, another part of the damping occurs inside the soft substrate itself due to the loss modulus of the PDMS gel. This latter contribution increases as the oscillation frequency increases and dominates at high frequencies. From the numerical evaluation of Eq. (14), and in the small $B$ case (*i.e.* $B < 2$, which is valid for the samples employed here), the limiting value of $G''$ at small distances ($D/D_c \to 0$) is found to depend on the moduli of the viscoelastic substrate via the relation:

$$\begin{cases} G' \simeq \frac{6\pi\eta R^2\omega}{D_c}(2.01 - 0.77\, B(\omega)), \\ G'' \simeq \frac{6\pi\eta R^2\omega}{D_c}(1.16 + 1.34\, B(\omega)). \end{cases} \qquad (16)$$

Moreover, comparing Figs. 3a and 3b, we observe that for the lower oscillation frequency, the real and imaginary parts of the mechanical impedance merge and saturate to the same value at small distances, whereas for the higher frequency the imaginary part remains larger than the real part at all distances. Furthermore, the imaginary part of the mechanical impedance increases as the frequency increases. The dissipation is thus more pronounced at large frequency.

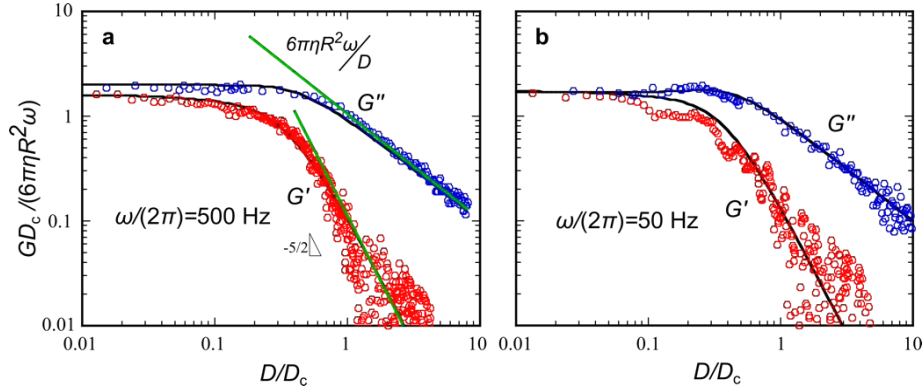

*Figure 3: Normalized real and imaginary parts of the mechanical impedance as functions of normalized distance, for two oscillation frequencies. The solid lines are fitting curves using Eq. (14). a) Results for an oscillation frequency of 500 Hz, as obtained from the data shown in Fig. 2b and Eq. (2). The extracted free parameters are the storage and loss moduli, $E' = 3.2 \pm 0.5$ kPa and $E'' = 2.0 \pm 0.2$ kPa, respectively. b) Results for an oscillation frequency of 50 Hz. The extracted moduli are $E' = 1.6 \pm 0.4$ kPa and $E'' = 0.69 \pm 0.10$ kPa.*

The extracted storage modulus $E'$ and loss modulus $E''$ for various oscillation frequencies are shown in Fig. 4. Both moduli increase as the frequency increases. In our experiment, the PDMS sample is a viscoelastic material, and the frequency dependence of its complex Young's modulus can be modeled by the *Chasset–Thirion* law [37,42,49,50]:

$$E(\omega) = E'(\omega) + jE''(\omega) = E_0(1 + (j\omega\tau)^n), \qquad (17)$$

where $E_0$ indicates the static ($\omega = 0$) Young's modulus, $\tau$ is the relaxation time and $n$ is an empirical exponent, which all depend on the sample preparation (*i.e.* curing procedure, stoichiometric ratio between the PDMS and curing agent).

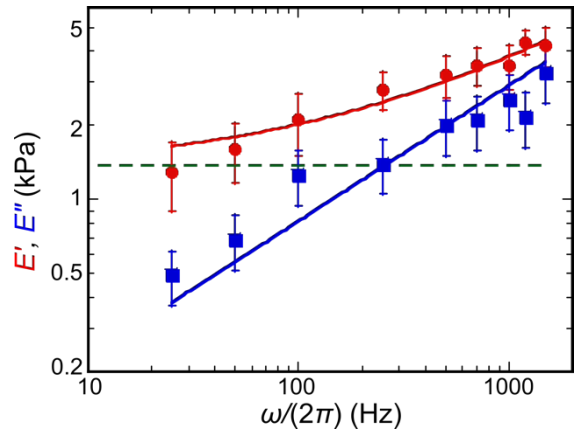

*Figure 4: Storage modulus $E'$(red) and loss modulus $E''$(blue) obtained from the fit in Fig.3 versus the frequency. The green dashed line represents the value of static modulus $E_0 = 1.3 \pm 0.1$ kPa measured from the independent indentation measurement. The solid lines are fits from Eq. (17), from which we get two free parameters: $n = 0.55 \pm 0.05$ and $\tau = 0.9 \pm 0.1$ ms.*

To characterize the static modulus $E_0$ of the substrate, the DC component of the cantilever's deflection versus the piezo displacement is also recorded, which allows us to construct the force-indentation curve. The static force $F$ is obtained by multiplying the DC component of the deflection by the spring constant $k_c$ of the cantilever. The indentation depth $\delta$ is obtained by subtracting the cantilever's deflection from the piezo displacement. The contact origin ($\delta = 0$) is defined as the position where the deflection increases sharply. The relation between the force and indentation is given by the well-known Hertz model:

$$F = \frac{16}{9} E_0 R^{\frac{1}{2}} \delta^{\frac{3}{2}}. \qquad (18)$$

Figure 5 shows the measured force $F$ as a function of the indentation depth $\delta$. The solid line represents the best fit from Eq. (18). From the fit, we obtain the static Young's modulus $E_0 = $ *1.3 ± 0.1* kPa as a single free parameter.

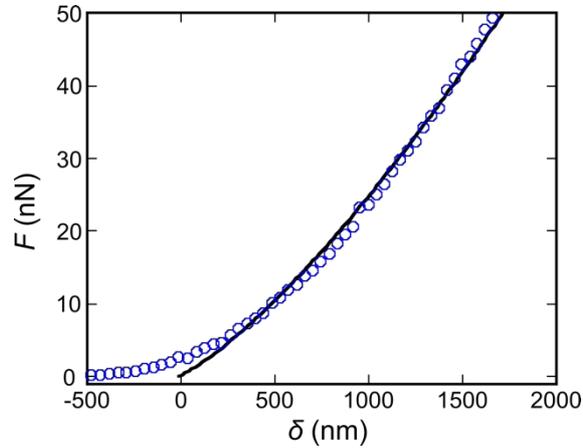

*Figure 5: Measured static force F versus indentation depth δ. The solid line shows the best fit to Eq. (18), from which we obtain the static Young's modulus $E_0 = 1.3 \pm 0.1$ kPa.*

With the obtained value of the static modulus $E_0$, the extracted storage and loss moduli are fitted by Eq. (17) where the relaxation time $\tau$ and the exponent $n$ are taken as fitting parameters. In Fig. 4, the solid lines are the best fits, from which we obtain: $n = $ *0.55 ± 0.05* and $\tau = $ *0.9 ± 0.1* ms. The empirical exponent $n$ typically takes values between *1/2* and *2/3* [51].

As a final remark, we recall that the PDMS layer is observed to swell and to increase in thickness by a factor ~5 when immersed in dodecane. Therefore, the PDMS layer contains a large amount of dodecane molecules that may flow across the pores of the network upon compression. Remarkably, the gel appears here to behave mechanically as a viscoelastic material following the *Chasset-Thirion* law, with no clear evidence of poroelastic response.

### *Conclusion:*

We have studied the viscoelastic response of crosslinked PDMS substrates at several frequencies, using a contactless AFM method. Specifically, a spherical colloidal probe attached to the AFM cantilever is vibrated in a liquid environment close and normally to a PDMS gel, in order to generate a nanoscale lubrication flow within the gap. This viscous flow generates a hydrodynamic pressure that deforms the soft substrate, leading to an elastohydrodynamic coupling. Based on soft-lubrication theory, we developed a model to calculate the resulting mechanical force exerted on the colloidal probe, including the viscoelasticity of the substrate. Using this model, the storage and loss elastic moduli of the PDMS gel were measured as functions of the frequency. The frequency dependencies of the storage and loss moduli were found to be in good agreement with the *Chasset–Thirion* law for viscoelastic solids. Altogether, this work demonstrates the robustness of broadband contactless AFM rheological methods. Such methods might be of interest for the gentle and precise frequency-dependent investigation of the mechanical behavior of thin, soft, fragile, immersed and/or alive systems, such as: polymers coatings, bubbles, biological membranes, etc.

**Acknowledgements:**

The authors thank Elisabeth Charlaix, Yacine Amarouchene and Nicolas Fares for fruitful discussions. They acknowledge financial support from the Agence Nationale de la Recherche (grants ANR-19-CE30-0012, ANR-21-ERCC-0010-01 *EMetBrown,* and ANR-21-CE06-0029 *Softer*) and the China Scholarship Council. They also thank the Soft Matter Collaborative Research Unit, Frontier Research Center for Advanced Material and Life Science, Faculty of Advanced Life Science at Hokkaido University, Sapporo, Japan.

**References:**
[1]	C. M. Stafford *et al.*, A buckling-based metrology for measuring the elastic moduli of polymeric thin films, Nature materials **3**, 545 (2004).
[2]	a. P S Alexopoulos and T. C. O'Sullivan, Mechanical Properties of Thin Films, Annual Review of Materials Science **20**, 391 (1990).
[3]	W. Allers, C. Hahn, M. Löhndorf, S. Lukas, S. Pan, U. Schwarz, and R. Wiesendanger, Nanomechanical investigations and modifications of thin films based on scanning force methods, Nanotechnology **7**, 346 (1996).
[4]	M. Kolle and S. Lee, Progress and opportunities in soft photonics and biologically inspired optics, Advanced Materials **30**, 1702669 (2018).
[5]	Z. Zhang *et al.*, in *Photonics* (Multidisciplinary Digital Publishing Institute, 2015), pp. 1005.
[6]	K. Xie, A. Glasser, S. Shinde, Z. Zhang, J. M. Rampnoux, A. Maali, E. Cloutet, G. Hadziioannou, and H. Kellay, Delamination and Wrinkling of Flexible Conductive Polymer Thin Films, Advanced Functional Materials, 2009039 (2021).
[7]	A. I. Hofmann, E. Cloutet, and G. Hadziioannou, Materials for transparent electrodes: from metal oxides to organic alternatives, Advanced Electronic Materials **4**, 1700412 (2018).
[8]	A. Mateescu, Y. Wang, J. Dostalek, and U. Jonas, Thin hydrogel films for optical biosensor applications, Membranes **2**, 40 (2012).
[9]	P. C. Jerónimo, A. N. Araújo, and M. C. B. Montenegro, Optical sensors and biosensors based on sol–gel films, Talanta **72**, 13 (2007).
[10]	A. Härtl *et al.*, Protein-modified nanocrystalline diamond thin films for biosensor applications, Nature Materials **3**, 736 (2004).
[11]	S. S. Asif, K. Wahl, and R. Colton, Nanoindentation and contact stiffness measurement using force modulation with a capacitive load-displacement transducer, Review of scientific instruments **70**, 2408 (1999).
[12]	P. McGuiggan, J. S. Wallace, D. T. Smith, I. Sridhar, Z. Zheng, and K. Johnson, Contact mechanics of layered elastic materials: experiment and theory, Journal of Physics D: Applied Physics **40**, 5984 (2007).
[13]	X. Chen and J. J. Vlassak, Numerical study on the measurement of thin film mechanical properties by means of nanoindentation, Journal of Materials Research **16**, 2974 (2001).

[14]  W. C. Oliver and G. M. Pharr, Measurement of hardness and elastic modulus by instrumented indentation: Advances in understanding and refinements to methodology, Journal of materials research **19**, 3 (2004).
[15]  R. García, R. Magerle, and R. Perez, Nanoscale compositional mapping with gentle forces, Nature Materials **6**, 405 (2007).
[16]  D. Wang and T. P. Russell, Advances in Atomic Force Microscopy for Probing Polymer Structure and Properties, Macromolecules **51**, 3 (2018).
[17]  Y. F. Dufrêne, T. Ando, R. Garcia, D. Alsteens, D. Martinez-Martin, A. Engel, C. Gerber, and D. J. Müller, Imaging modes of atomic force microscopy for application in molecular and cell biology, Nature Nanotechnology **12**, 295 (2017).
[18]  R. Garcia, Nanomechanical mapping of soft materials with the atomic force microscope: methods, theory and applications, Chemical Society Reviews **49**, 5850 (2020).
[19]  M. Heuberger, G. Dietler, and L. Schlapbach, Mapping the local Young's modulus by analysis of the elastic deformations occurring in atomic force microscopy, Nanotechnology **6**, 12 (1995).
[20]  M. R. VanLandingham, J. S. Villarrubia, W. F. Guthrie, and G. F. Meyers, Nanoindentation of polymers: an overview, Macromolecular Symposia **167**, 15 (2001).
[21]  H.-J. Butt, B. Cappella, and M. Kappl, Force measurements with the atomic force microscope: Technique, interpretation and applications, Surface Science Reports **59**, 1 (2005).
[22]  E. K. Dimitriadis, F. Horkay, J. Maresca, B. Kachar, and R. S. Chadwick, Determination of elastic moduli of thin layers of soft material using the atomic force microscope, Biophysical journal **82**, 2798 (2002).
[23]  J. T. Pham, F. Schellenberger, M. Kappl, and H.-J. Butt, From elasticity to capillarity in soft materials indentation, Physical Review Materials **1**, 015602 (2017).
[24]  K. L. Johnson, K. Kendall, and a. Roberts, Surface energy and the contact of elastic solids, Proceedings of the royal society of London. A. mathematical and physical sciences **324**, 301 (1971).
[25]  K. Johnson, Contact mechanics cambridge university press london, UK  (1985).
[26]  H.-J. Butt, J. T. Pham, and M. Kappl, Forces between a stiff and a soft surface, Current Opinion in Colloid & Interface Science **27**, 82 (2017).
[27]  E. Barthel and A. Perriot, Adhesive contact to a coated elastic substrate, Journal of Physics D: Applied Physics **40**, 1059 (2007).
[28]  K. R. Shull, Contact mechanics and the adhesion of soft solids, Materials Science and Engineering: R: Reports **36**, 1 (2002).
[29]  Y.-S. Chu, S. Dufour, J. P. Thiery, E. Perez, and F. Pincet, Johnson-Kendall-Roberts Theory Applied to Living Cells, Physical Review Letters **94**, 028102 (2005).
[30]  F. Kaveh, J. Ally, M. Kappl, and H.-J. r. Butt, Hydrodynamic force between a sphere and a soft, elastic surface, Langmuir **30**, 11619 (2014).
[31]  L. Garcia, C. Barraud, C. Picard, J. Giraud, E. Charlaix, and B. Cross, A micro-nano-rheometer for the mechanics of soft matter at interfaces, Review of Scientific Instruments **87**, 113906 (2016).
[32]  S. Leroy, A. Steinberger, C. Cottin-Bizonne, F. Restagno, L. Léger, and É. Charlaix, Hydrodynamic interaction between a spherical particle and an elastic surface: a gentle probe for soft thin films, Physical review letters **108**, 264501 (2012).
[33]  R. Villey, E. Martinot, C. Cottin-Bizonne, M. Phaner-Goutorbe, L. Léger, F. Restagno, and E. Charlaix, Effect of surface elasticity on the rheology of nanometric liquids, Physical review letters **111**, 215701 (2013).
[34]  D. Guan, C. Barraud, E. Charlaix, and P. Tong, Noncontact Viscoelastic Measurement of Polymer Thin Films in a Liquid Medium Using Long-Needle Atomic Force Microscopy, Langmuir **33**, 1385 (2017).
[35]  D. Guan, E. Charlaix, R. Z. Qi, and P. Tong, Noncontact viscoelastic imaging of living cells using a long-needle atomic force microscope with dual-frequency modulation, Physical Review Applied **8**, 044010 (2017).
[36]  J. Vega, A. Santamaria, A. Munoz-Escalona, and P. Lafuente, Small-amplitude oscillatory shear flow measurements as a tool to detect very low amounts of long chain branching in polyethylenes, Macromolecules **31**, 3639 (1998).

[37] M. Zhao, J. Dervaux, T. Narita, F. Lequeux, L. Limat, and M. Roché, Geometrical control of dissipation during the spreading of liquids on soft solids, Proceedings of the National Academy of Sciences **115**, 1748 (2018).

[38] M. Gardel, J. H. Shin, F. MacKintosh, L. Mahadevan, P. Matsudaira, and D. A. Weitz, Elastic behavior of cross-linked and bundled actin networks, Science **304**, 1301 (2004).

[39] R. H. Ewoldt, A. Hosoi, and G. H. McKinley, New measures for characterizing nonlinear viscoelasticity in large amplitude oscillatory shear, Journal of Rheology **52**, 1427 (2008).

[40] H. T. Banks, S. Hu, and Z. R. Kenz, A brief review of elasticity and viscoelasticity for solids, Advances in Applied Mathematics and Mechanics **3**, 1 (2011).

[41] D. T. Chen, Q. Wen, P. A. Janmey, J. C. Crocker, and A. G. Yodh, Rheology of soft materials, Annu. Rev. Condens. Matter Phys. **1**, 301 (2010).

[42] E. Rolley, J. H. Snoeijer, and B. Andreotti, A flexible rheometer design to measure the visco-elastic response of soft solids over a wide range of frequency, Review of scientific instruments **90**, 023906 (2019).

[43] A. Maali and R. Boisgard, Precise damping and stiffness extraction in acoustic driven cantilever in liquid, Journal of Applied Physics **114**, 144302 (2013).

[44] V. S. Craig and C. Neto, In situ calibration of colloid probe cantilevers in force microscopy: hydrodynamic drag on a sphere approaching a wall, Langmuir **17**, 6018 (2001).

[45] C. Jai, T. Cohen-Bouhacina, and A. Maali, Analytical description of the motion of an acoustic-driven atomic force microscope cantilever in liquid, Applied physics letters **90**, 113512 (2007).

[46] A. Maali, R. Boisgard, H. Chraibi, Z. Zhang, H. Kellay, and A. Würger, Viscoelastic drag forces and crossover from no-slip to slip boundary conditions for flow near air-water interfaces, Physical review letters **118**, 084501 (2017).

[47] S. Leroy and E. Charlaix, Hydrodynamic interactions for the measurement of thin film elastic properties, Journal of Fluid Mechanics **674**, 389 (2011).

[48] E. Gacoin, C. Fretigny, A. Chateauminois, A. Perriot, and E. Barthel, Measurement of the mechanical properties of thin films mechanically confined within contacts, Tribology Letters **21**, 245 (2006).

[49] R. Chasset and P. Thirion, in *Proceedings of the International Conference* (Prins JA Ed. North-Holland Publishing Co., Amsterdam, 1965), pp. 345.

[50] D. C. Agudelo, L. E. Roth, D. A. Vega, E. M. Vallés, and M. A. Villar, Dynamic response of transiently trapped entanglements in polymer networks, Polymer **55**, 1061 (2014).

[51] J. C. Scanlan and H. H. Winter, Composition dependence of the viscoelasticity of end-linked poly (dimethylsiloxane) at the gel point, Macromolecules **24**, 47 (1991).